\documentclass{iopart}
\usepackage{iopams}
\expandafter\let\csname equation*\endcsname\relax
\expandafter\let\csname endequation*\endcsname\relax

\usepackage{appendix}
\usepackage{fancyvrb}
\usepackage{url}
\usepackage{hyperref}
\usepackage{geometry}
\usepackage{enumerate}

\usepackage{mathtools,bm}
\usepackage{amsmath,amsfonts,amssymb}   
\usepackage{amsmath}
\usepackage{amssymb}

\usepackage{float}
\usepackage{graphicx}
\usepackage{caption,subcaption}
\usepackage{tikz}
\usepackage{pgfplots}
\pgfplotsset{compat=newest}

\usetikzlibrary{patterns, arrows.meta, calc}
\usepgfplotslibrary{fillbetween}

\usepackage{xcolor}
\definecolor{lightblue}{rgb}{0.0,0.5,0.8}
\definecolor{HMSRed}{HTML}{C5112E}  
\definecolor{MGHBlue}{HTML}{008BB0} 
\definecolor{MGHGrey}{HTML}{626365} 

\newcommand{\quotes}[1]{``#1''}   

\begin{document}

\title[Dynamic MLC sequencing with splines ]{Dynamic fluence map sequencing using piecewise linear leaf position functions}
\author{Matthew Kelly$^1$, Jacobus H.M. van Amerongen$^2$, Marleen Balvert$^{2,3,4}$, David Craft$^5$}
\address{$^1$ Department of Mechanical Engineering, Tufts University, Medford MA 02155, USA}
\address{$^2$ Department of Econometrics and Operations Research/Center for Economic Research (CentER), Tilburg University, PO Box 90153, 5000 LE Tilburg, The Netherlands}
\address{$^3$ Centrum Wiskunde \& Informatica (CWI), P.O. Box 94079, 1090 GB Amsterdam, The Netherlands}
\address{$^4$ Department of Biology, University of Utrecht, Padualaan 8, 3584 CH Utrecht, The Netherlands}
\address{$^5$ Department of Radiation Oncology, Massachusetts General Hospital and Harvard Medical School, Boston, MA 02114, USA}
\ead{dcraft@mgh.harvard.edu}

\begin{abstract}
  Within the setting of intensity modulated radiation therapy (IMRT) and the fully continuous version of IMRT called volumetric modulated radiation therapy (VMAT), we consider the problem of matching a given fluence map as well as possible in limited time by the use of a linear accelerator (linac) with a multi-leaf collimator (MLC). We introduce two modeling strategies to manage the nonconvexity and the associated local minima of this problem. The first is the use of linear splines to model the MLC leaf positions as functions of time. The second is a progressively controllable smooth model (instead of a step function) of how the leaves block the fluence radiation. We propose a two part solution: an outer loop that optimizes the dose rate pattern over time, and an inner loop that given a dose rate pattern optimizes the leaf trajectories.
\end{abstract}

\section{Introduction}
The optimal dynamic delivery of a given fluence map by a multi-leaf collimator (MLC) remains a difficult, unsolved problem.
The sliding-window leaf-sweep algorithm (SWLS) \cite{leafsweep},
    in which the MLC leaves cross the treatment field in a unidirectional fashion,
    achieves perfect fluence map replication if sufficient time is available \cite{Stein94}.
However, the SWLS algorithm is not in general efficient with respect to the required delivery time \cite{balvertcraft}.
Time is an important aspect of VMAT and IMRT treatment plans, for several reasons:
\begin{enumerate}[i)]
  \item Shorter treatments allow the treatment facility to help more patients on a given set of radiation therapy machines,
        which is particularly relevant to developing countries as these machines are expensive.
  \item The effect of patient movement on delivery inaccuracy increases in the time the patient is exposed to radiation.
  \item In general, there is a trade-off between dose quality and delivery time, and given how widespread the use of radiation therapy is in treating cancer, it makes sense to put in effort to assure that we are on the Pareto optimal frontier regarding these two conflicting objectives.

\end{enumerate}
Several studies have investigated the trade-off between treatment time and plan quality \cite{tradeoffSalari,tradeoffMCO,tradeoffCraft,balvertcraft}.
\cite{balvertcraft} were the first to include treatment time directly in a dynamic leaf sequencing step of the treatment plan optimization.
They constructed the trade-off curve between delivery time and fluence map matching accuracy by optimizing leaf trajectories and dose rate patterns for a sequence of delivery times.
For a given fluence map and fixed delivery time, the challenge of optimizing the leaf trajectories and dose rate so that the given fluence map is matched as accurately as possible, subject to machine restrictions, presents a high dimensional nonconvex optimization problem.
The nonconvexity of the fluence map matching problem leads to a large number of local minima.
For a thorough introduction to the complexities of dynamic fluence map delivery (which generally arises in the context of dynamic IMRT and VMAT), see \cite{balvertcraft} and \cite{unkvmatreview}. 
Briefly, we note that there are two broad solution types available in commercial systems: sliding window derived approaches, which use smaller MLC aperture openings and thus lead to longer delivery times, and step-and-shoot derived approaches. Neither of these approaches can guarantee optimality of the solution: sliding window does not allow for bidirectional leaf motions, which may be necessary for optimality \cite{balvertcraft}, and step-and-shoot is not designed for continuous delivery. It is thus important to attack the problem in a more general setting: broadly searching over valid leaf trajectories and dose rates to determine the optimal delivery pattern without restricting to one of the above settings. This was, to the best of our knowledge, only done in \cite{balvertcraft} and \cite{thesisKvA}, where continuous leaf motions were represented by a discretized motion in the optimization, leading to a gap between the optimized and the delivered plan.  Due to the rising clinical usage of VMAT, which is by nature continuous, we believe it is prudent to continue the basic research on optimal dynamic fluence delivery in search of a clinically usable approach for continuous leaf motions.

In this report, we present a new approach for optimizing the continuous leaf motion dynamics to match a given fluence map. A logical way to search for a combined dose rate pattern and leaf motion dynamics to best produce a given fluence map is to do a nested optimization with the dose rate search in the outer loop and the leaf trajectory search in the inner loop \cite{thesisKvA}. The rationale for this is that once the outer loop sets a dose rate profile, the MLC leaf pairs can be optimized independently (setting a dose rate profile decouples the MLC leaf rows) \cite{balvertcraft,thesisKvA}. We only consider the inner search for optimal leaf trajectories and hence assume a dose rate pattern is given. Although our method can be applied to an arbitrary dose rate profile, we use a constant dose rate.

\section{Methods}
\label{sec:model}

Our starting point is a fluence map $m$ that has been optimized, along with additional fluence maps located at given angles around the patient, to collectively yield a dose distribution optimized for the particular patient's geometry (location of tumor and all nearby organs) and dose prescription. The algorithms set forth in this paper determine how to construct a single given fluence map by moving the leaves of the MLC within the field, for a given dose rate pattern. Our optimization allows the leaves to move back and forth, a requirement for achieving optimal motions in the setting of a general (non-constant) dose rate, as shown in the Appendix of \cite{balvertcraft}. Moreover, we allow the leaves of every pair to start and end at arbitrary feasible locations within the field, not necessarily at the bounds of the treatment field, as these restrictions can also be suboptimal \cite{thesisKvA}. Thus the problem we model and solve is the dynamic IMRT field delivery problem, which is a subproblem of the full dynamic VMAT problem \cite{vmerge}.

We assume the fluence map $m$ is given as a matrix where the rows correspond to the leaf pairs, and the columns are the discretely optimized fluence bixels across the field, the latter of which can be as finely discretized as one wishes. Typical length scales are on the order of 0.5 cm for both the row height of the MLC leaves and the across-the-row discretization.

Let $x^i_L(t)$ and $x^i_R(t)$ denote the leaf position of the $i$th left and right leaves respectively, at time $t$. For a fixed dose rate pattern, the leaf rows can be optimized independently (neglecting the small coupling terms created by the tongue-and-groove mechanism on the real machine, and output factor considerations, see \cite{unkvmatreview}), so for the remainder of the leaf motion algorithm development, we consider only a single leaf row, and therefore drop the $i$ superscript. Let $f(x)$ be the target fluence that should be delivered for that row. Note if $f$ is obtained from an optimized fluence map $m$ it is piecewise constant, but in general $f$ can also be smooth. We assume the total allowed treatment delivery time $T$ is given. Our goal is to compute the leaf trajectories $x_L(t)$ and $x_R(t)$  to recreate the fluence row $f(x)$ as best as possible, while accounting for maximum leaf speed and collision constraints. 

\begin{figure}[htp]
\centering
    \begin{tikzpicture}[remember picture]
        \pgfplotsset{holdot/.style={color=black,only marks,mark=*,mark size=1.5pt}}
        \pgfplotsset{soldot/.style={color=white,only marks,mark=*,mark size=1.5pt}}
        \begin{axis}[
            xmin=0, xmax=10.50,
            ymin=0, ymax=4.5,
            grid = both,
            grid style = dotted,
            axis x line = bottom,
            axis y line = left,
            enlargelimits = {abs=0.0},
            axis line style = {-Latex[round]},
            yticklabels = {,$x$,},
            ytick = {0.2,2,4},
            xtick = \empty,
            ylabel = {\scriptsize{position (cm)}},
            axis equal image
        ]

            \node[] (traj-1) at (axis cs:1,4.2) {};
            \node[] (traj-2) at (axis cs:1.88,4.2) {};
            \node[] (traj-3) at (axis cs:3.5,4.2) {};
            \node[] (traj-4) at (axis cs:4.17,4.2) {};
            \node[] (traj-5) at (axis cs:5.2,4.2) {};
            \node[] (traj-6) at (axis cs:5.85,4.2) {};
            \node[] (traj-10) at (axis cs:10,4.2) {};

            \draw[MGHBlue] (1,2) -- (1.89,2);
            \draw[MGHBlue] (3.5,2) -- (4.17,2.0);
            \draw[MGHBlue] (5.2,2) -- (5.85,2);

            \addplot[name path global = leafl, black, smooth, tension=0.75] coordinates{(0,0.2)(1,2)(2.5,3.7)(4.7,1.9)(8.5,4)(10,4)};
            \addplot[name path global = leafr, black, smooth, tension=0.75] coordinates{(0,0.2)(1,0.5)(2.5,2.5)(4.5,1.4)(6.5,2.3)(8,2.1)(10,4)};

            \node[] at (axis cs:6.8,3.7) {$x_L$};
            \node[] at (axis cs:8.2,1.5) {$x_R$};
            \draw[MGHGrey!50] (7.2,3.7) -- (7.6,3.5);
            \draw[MGHGrey!50] (8.2,1.8) -- (8,2.1);

        \end{axis}
    \end{tikzpicture}

    \begin{tikzpicture}[remember picture]
        \pgfplotsset{holdot/.style={color=black,only marks,mark=*,mark size=1.5pt}}
        \pgfplotsset{soldot/.style={color=white,only marks,mark=*,mark size=1.5pt}}
        \begin{axis}[
            xmin=0, xmax=10.50,
            ymin=0, ymax=3.0,
            ymajorgrids = true,
            grid style = dotted,
            axis x line = bottom,
            axis y line = left,
            enlargelimits = {abs=0.0},
            axis line style = {-Latex[round]},
            xticklabels = {0, $T$},
            xtick = {0,10},
            yticklabels = {$d$},
            ytick = {2.5},
            xlabel style = {at={(ticklabel* cs:0.5,1.5)},anchor=north},
            xlabel = {\scriptsize{time(s)}},
            ylabel = {\scriptsize{dose rate (MU/s)}},
            axis equal image
        ]

            \node[] (dose-1) at (axis cs:1,-0.2) {};
            \node[] (dose-2) at (axis cs:1.88,-0.2) {};
            \node[] (dose-3) at (axis cs:3.5,-0.2) {};
            \node[] (dose-4) at (axis cs:4.17,-0.2) {};
            \node[] (dose-5) at (axis cs:5.2,-0.2) {};
            \node[] (dose-6) at (axis cs:5.85,-0.2) {};
            \node[] (dose-10) at (axis cs:10,-0.2) {};

            \addplot[name path global = dose, black, smooth, tension=0.75] coordinates{(0,1.25)(2.5,2.4)(6,0.4)(8,1.9)(10,1.7)};\
            \addplot[name path global = xaxis, draw=none, domain=0:11] {0};

            \addplot [thick, color=blue, fill=MGHGrey, fill opacity=0.2]
                fill between[of=xaxis and dose, soft clip={domain=1:1.88}];
            \addplot [thick, color=blue, fill=MGHGrey, fill opacity=0.2]
                fill between[of=xaxis and dose, soft clip={domain=3.5:4.17}];
            \addplot [thick, color=blue, fill=MGHGrey, fill opacity=0.2]
                fill between[of=dose and xaxis, soft clip={domain=5.2:5.85}];

            \addplot[fill = white]
                fill between[of=xaxis and dose, soft clip={domain=1.88:2}]; 
            \addplot[fill = white]
                fill between[of=xaxis and dose, soft clip={domain=4.17:4.5}]; 
            \addplot[fill = white]
                fill between[of=xaxis and dose, soft clip={domain=5.85:6}]; 
             \draw[black] (0,0) -- (10,0);

        \end{axis}
    \end{tikzpicture}

    \begin{tikzpicture}[remember picture,overlay]
        \draw[dashed,MGHGrey!60] (traj-1) -- (dose-1);
        \draw[dashed,MGHGrey!60] (traj-2) -- (dose-2);
        \draw[dashed,MGHGrey!60] (traj-3) -- (dose-3);
        \draw[dashed,MGHGrey!60] (traj-4) -- (dose-4);
        \draw[dashed,MGHGrey!60] (traj-5) -- (dose-5);
        \draw[dashed,MGHGrey!60] (traj-6) -- (dose-6);
        \draw[dashed,MGHGrey!60] (traj-10) -- (dose-10);
    \end{tikzpicture}

\vspace{-0.7cm}
\caption[Illustration of administered fluence]{
    Illustration of administered fluence.
    In the upper panel, the upper and lower lines display the trajectories of the left and right leaves, respectively; the lower panel shows the dose rate pattern.
    The dose administered to a position $x$, $g(x)$, equals the integral (shaded area) of the dose rate $d(t)$ over the moments in time $\mathcal{T}(x)$ (blue lines) that position is exposed.
    }
\label{fig:administeredDose}
\end{figure}
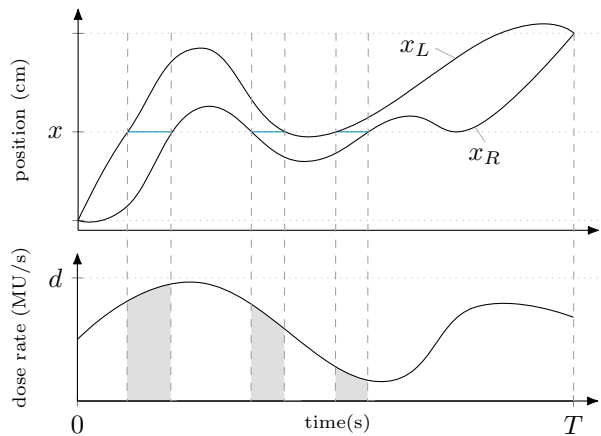

The fluence achieved at each position $x$ is $g(x)$, which is the time-integral of the dose rate for the times that this position is exposed to the radiation source.
The time domain of exposure $\mathcal{T}(x)$ is the set of times (in general a disconnected set) when the position $x$ is not blocked by either of the leaves, i.e.,
$\mathcal{T}(x)$ is the set of all times $t$ such that $x_L(t) \le x \leq x_R(t)$,

\begin{equation}
g(x) = \int_{t \in \mathcal{T}(x)} \! d(t) \, dt ,
\label{eqn:deliveredFluenceDose}
\end{equation}
as illustrated by Figure \ref{fig:administeredDose}.
The full fluence map matching problem, including the dose rate search, can be stated as follows. Find the leaf trajectories $x_L(t)$ and $x_R(t)$ and dose rate pattern $d(t)$
that minimize the squared integral error between the target fluence $f(x)$ and the delivered fluence $g(x)$:

\begin{equation}
\underset{d(t), \, x_L(t), \, x_R(t)}{\operatorname{argmin}}
\int_X \bigg(f(x) - g(x)\bigg)^2 dx 
\label{eqn:fluenceMapOptimization}
\end{equation}
\noindent subject to feasibility constraints on the dose rate and leaf trajectory functions.

Next we describe the method that we use to convert this mathematical optimization problem into a format that can be solved using standard nonlinear programming (NLP) solvers such as FMINCON \cite{MatlabOptimizationToolbox2014}, SNOPT \cite{Snopt7}, or IPOPT \cite{Wachter2006}.

The first step in this process is to select the computational representation for the leaf trajectories, for which we use piecewise linear functions. The second step is to compute the integrals in the objective function using methods that are smooth and consistent, a critical step for obtaining good results from the NLP solver \cite{Betts2010}.

\subsection{Spline Representation}

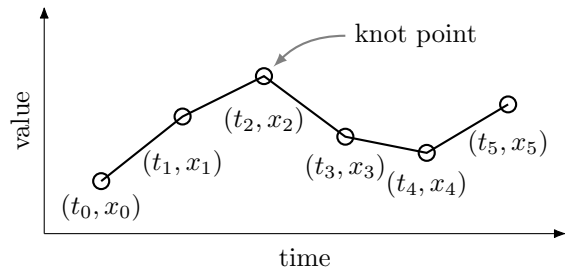
\begin{figure}[htp]
\centering
\begin{tikzpicture}
    \begin{axis}[
        xmin=0, xmax=6.0,
        ymin=0, ymax=2.4,
        axis x line = bottom,
        axis y line = left,
        enlargelimits = {abs=0.2},
        axis line style = {-Latex[round]},
        ytick = \empty,
        xtick = \empty,
        ylabel = {value},
        xlabel = {time},
        axis equal image
        ]
        
    \addplot[   thick,
        color=black,
        solid,
        mark=o,
        mark size=3,
        mark options={solid},
        visualization depends on=\thisrow{alignment} \as \alignment,
        nodes near coords, 
        point meta=explicit symbolic, 
        every node near coord/.style={yshift=\alignment, anchor=north} 
        ] table [
            meta index=2 
        ] {
        x       y       label       alignment
        0.5     0.45    $(t_0,x_0)$       -1
        1.5     1.25    $(t_1,x_1)$       -9
        2.5     1.75    $(t_2,x_2)$       -8
        3.5     1.0     $(t_3,x_3)$       -3
        4.5     0.8     $(t_4,x_4)$       -3
        5.5     1.4     $(t_5,x_5)$       -6
        };

        \node at (2.5,1.75) (knot) {};
        \node[anchor=west] at (3.5,2.25) (description) {knot point};
        \draw[color=gray, thick] (description) edge[out=180,in=45,-latex] (knot);

    \end{axis}
\end{tikzpicture}

\caption[Linear Spline]{ 
    Leaf position trajectories are represented using linear splines. 
}
\label{fig:linearSpline}
\end{figure}

There are two continuous functions, the position of the left and right leaves, $x_L(t)$ and $x_R(t)$, that must be computed by the optimization (note, if we were including the dose rate in our optimization there would be three functions to optimize). We use piecewise linear functions (linear splines) to represent these. A linear spline is fully defined by its value at the knot points $t_k$: $x_{L,k}$, $x_{R,k}$. An example of a linear spline is shown in Figure \ref{fig:linearSpline}.


\subsection{Integral Computation with Blocking Function k()}
\label{sec:IntegralComputationWithBlockingFunction}

There are two issues with computing the integral in Equation \ref{eqn:fluenceMapOptimization} directly: 1) computing the domain $\mathcal{T}(x)$ requires a root solve (or inverting the leaf trajectories), and 2) the domain of $\mathcal{T}(x)$ can change from being simply connected to discontinuous during an optimization. Both of these issues would likely cause convergence failures in the NLP solver, in part by causing a change in the sparsity pattern of the gradient
between successive iterations.

Our first step is to rewrite the integral using a blocking function $k(t,x)$, which has a value of one when the leaves at time $t$ are passing radiation at location $x$ and zero when the leaves are blocking radiation. This allows us to rewrite the integral using the constant bounds $[0, T]$:

\begin{equation}
  g(x) = \int_{t=0}^T \! k(t, x) \cdot d(t) \, dt .
  \label{eqn:fluenceDoseSimpleBounds}
\end{equation}

We now have a scalar integral and we can use any standard quadrature method to evaluate (\ref{eqn:fluenceDoseSimpleBounds}).
In our case we use the midpoint (rectangle) quadrature rule.

As just defined, our fluence blocking function $k(t,x)$
would also have a discontinuous gradient, which would cause convergence issues in the optimization.
Therefore, we use an exponential sigmoid function 
to approximate the step changes in the blocking function, where $\alpha$ is the smoothing parameter: 
\begin{equation}
  s(x, \alpha) = (1 + e^{-\alpha x})^{-1} .
  \label{eqn:sigmoidEquation}
\end{equation}
A small value of $\alpha$ corresponds to heavy smoothing and faster convergence in the optimization, while a large value of $\alpha$ will provide a more accurate model at the expense of a more difficult optimization.
We can then combine the smoothing function for each leaf to get the combined blocking function:
\begin{equation}
  k(t, x) \approx \sqrt{s\big(x_R(t) -x, \, \alpha\big) \; \cdot \; s\big(x -x_L(t), \, \alpha\big)} .
  \label{eqn:blockingFunction}
\end{equation}
\noindent In practice it is useful to define the $\alpha$ parameter in terms of 
a smoothing distance $\Delta x$ and 
the fraction $\gamma$ that the blocking function changed over that distance.
For example, $\Delta x = 0.05$ cm and $\gamma = 0.98$ means that 
the blocking function changes from $0.01$ to $0.99$ over a distance of $0.05$ cm. $\alpha$ can then be computed as:

\begin{equation}
  \alpha = \frac{-2}{\Delta x} \; \ln \! \left( \frac{1 - \gamma}{1 + \gamma} \right) .
  \label{eqn:SmoothingDistanceParameter}
\end{equation}

\begin{figure}
  \centering
  \includegraphics[width=\textwidth]{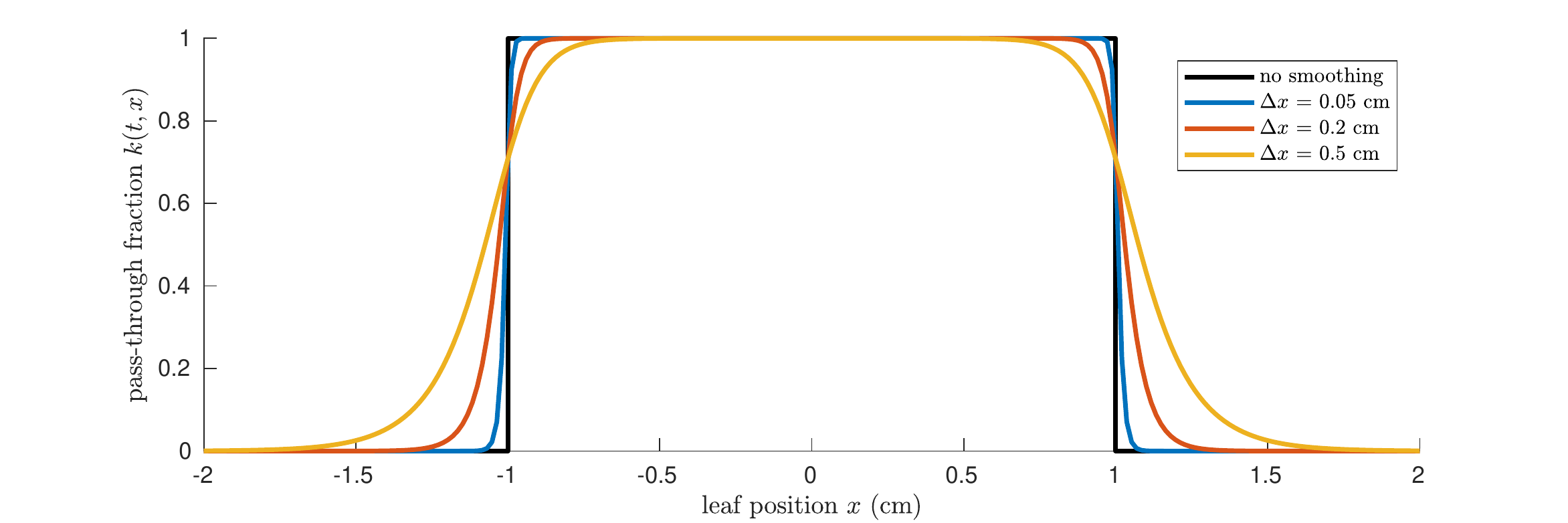}
  \caption{Visualization of smoothing parameters in the blocking function (\ref{eqn:blockingFunction}). The right and left leaves are at $x_R = 1$ cm and $x_L = -1$ cm respectively. The solid black line shows the case without smoothing, and the remaining lines show light smoothing ($\Delta x = 0.05$ cm), moderate smoothing ($\Delta x = 0.2$ cm), and heavy smoothing ($\Delta x = 0.5$ cm). In each of these three cases we use a value of $\gamma = 0.95$.}
  \label{fig:visualizeExponentialSmoothing}
\end{figure}

Figure \ref{fig:visualizeExponentialSmoothing} shows three values of the smoothing parameter for the blocking function $k(t, x)$, where $x_R = 1$ and $x_L = -1$, and compares the function to the case without smoothing.

\subsection{Computing Leaf Trajectories as a Nonlinear Program}
\label{sec:LeafTrajectoryAsNLP}

The integral squared error objective function in formulation (\ref{eqn:fluenceMapOptimization}) is discretized for the numerical optimization procedure. We break the domain $[x_\text{min}, x_\text{max}]$ into $N_\text{fit}$ equal-width segments, and evaluate the fluence target and delivered fluence at the midpoint $x_k$ of each segment:
\begin{equation}
  \int_{x_\text{min}}^{x_\text{max}} \! \bigg( f(x) - g(x) \bigg)^2 \,dx  \approx \frac{x_\text{max} - x_\text{min}}{N_\text{fit}}
  \sum_{k = 1}^{N_\text{fit}} \! \bigg( f(x_k) - g(x_k) \bigg)^2.
  \label{eqn:discreteFittingObjective}
\end{equation}

Constraints that the leaves remain within the physical bounds of the fluence field and do not collide are given by:
\begin{equation}
  x_\text{min} \leq x_{L, k}
  \quad \quad
  x_{R, k} \leq x_\text{max}
  \quad \quad
  x_{L, k} \leq x_{R, k}
  \quad \quad
  \forall k .
  \label{eqn:PositionLimits}
\end{equation}
\noindent where leaf position is given by linear interpolation between the knot points. 

Velocity constraints can also be handled with linear inequalities:
\begin{equation}
  -v_\text{max} \leq \dot{x}_{L, k} \leq v_\text{max}
  \quad \quad
  -v_\text{max} \leq \dot{x}_{R, k} \leq v_\text{max}
  \quad \quad \forall k .
  \label{eqn:VelocityLimits}
\end{equation}
\noindent  where the velocity of each leaf on each spline segment is constant and given by:
\begin{equation}
  \dot{x}_{L, k} = \frac{x_{L, k+1} - x_{L, k}}{h_k}
  \quad \quad
  \dot{x}_{R, k} = \frac{x_{R, k+1} - x_{R, k}}{h_k},
\end{equation}
\noindent and $h$ is the distance between two knot points. 

\subsection{Iterative refinement of smoothing parameter}

The performance of the optimization, based on solve-time and accuracy, is highly dependent on the value of the smoothing parameter $\alpha$. With heavy smoothing the optimization will quickly converge to a \quotes{good} solution, but the smoothing distorts the objective function to the point where it is inaccurate. Conversely, with light smoothing (or no smoothing) the gradients in the optimization change quickly and the solver easily gets stuck in local minima and sometimes fails to converge.

This dependency on smoothing is common in trajectory optimization and there is a well known solution: iterative refinement. The idea is to initially solve the optimization using heavy smoothing, which gives a solution that is somewhat close to the true optimal solution. Then the optimization is solved again using the previous solution as the initial guess and with a smaller value of the smoothing parameter. This process is continued until the error in the objective function decreases to an acceptable level \cite{Srinivasan2006}.

\section{Results}
\begin{figure}
  \centering
  \includegraphics[]{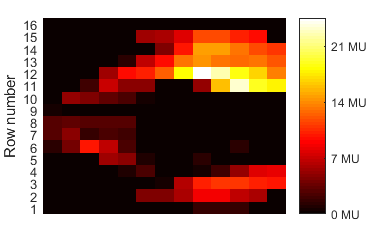}
  \caption{Target fluence map with 1 cm bixels, from the CORT dataset \cite{CORT14}.}
  \label{fig:targetmap}
\end{figure}

The method is demonstrated using a fluence map that is generated for a prostate patient with lymph nodes publicly available via the CORT dataset, see Figure \ref{fig:targetmap} \cite{CORT14}.
The bixel width is 1 cm.
First, we demonstrate the inner loop search using two fluence profiles, one with a bimodal and one with a unimodal shape which correspond to the 11$^\text{th}$ and 12$^\text{th}$ rows of this fluence map respectively.
These fluence profiles are depicted in Figure \ref{fig:sps_umt_prof} respectively Figure \ref{fig:sps_bmt_prof}.
Next, we demonstrate the overall method on the entire fluence map depicted in Figure \ref{fig:targetmap}.
We set the dose rate level to its maximum level thus not performing the outer loop search where dose rates are optimized. In what would be the inner loop search we solve the leaf trajectory optimization problem for each row of the entire (near-unimodal) fluence map and its transposed (near-bimodal) version (due to row independence, this search can be done in parallel, as described above). 

We assume a maximum leaf speed of 3 cm/s and a dose rate of 10 MU/s. 
The performance is evaluated based on the fluence profile (fluence map) matching quality of the final solution, measured by the sum of squared integral errors (Equation \ref{eqn:fluenceMapOptimization}) over all leaf rows considered, and the CPU time the algorithm needs to compute the solution.
Computations are performed in Matlab (R2017a) on a desktop computer with a 3.4GHz quad-core Intel i5-3570K processor.

All of the experiments in this report use two segments per second for the leaf trajectoriy splines. This number was chosen using a pilot study.
If fewer knot points were used,  then the ability to fit arbitrary fluence profiles was diminished.
If more knot points were used then there was a minor improvement in fitting, but an increase in computation time and it was more difficult to find a viable smoothing schedule. 


\subsection{Smoothing Parameter Schedule}
\begin{figure*}[t!]
    \centering
    \begin{subfigure}[b]{0.5\textwidth}
        \centering
        \includegraphics[width=\textwidth]{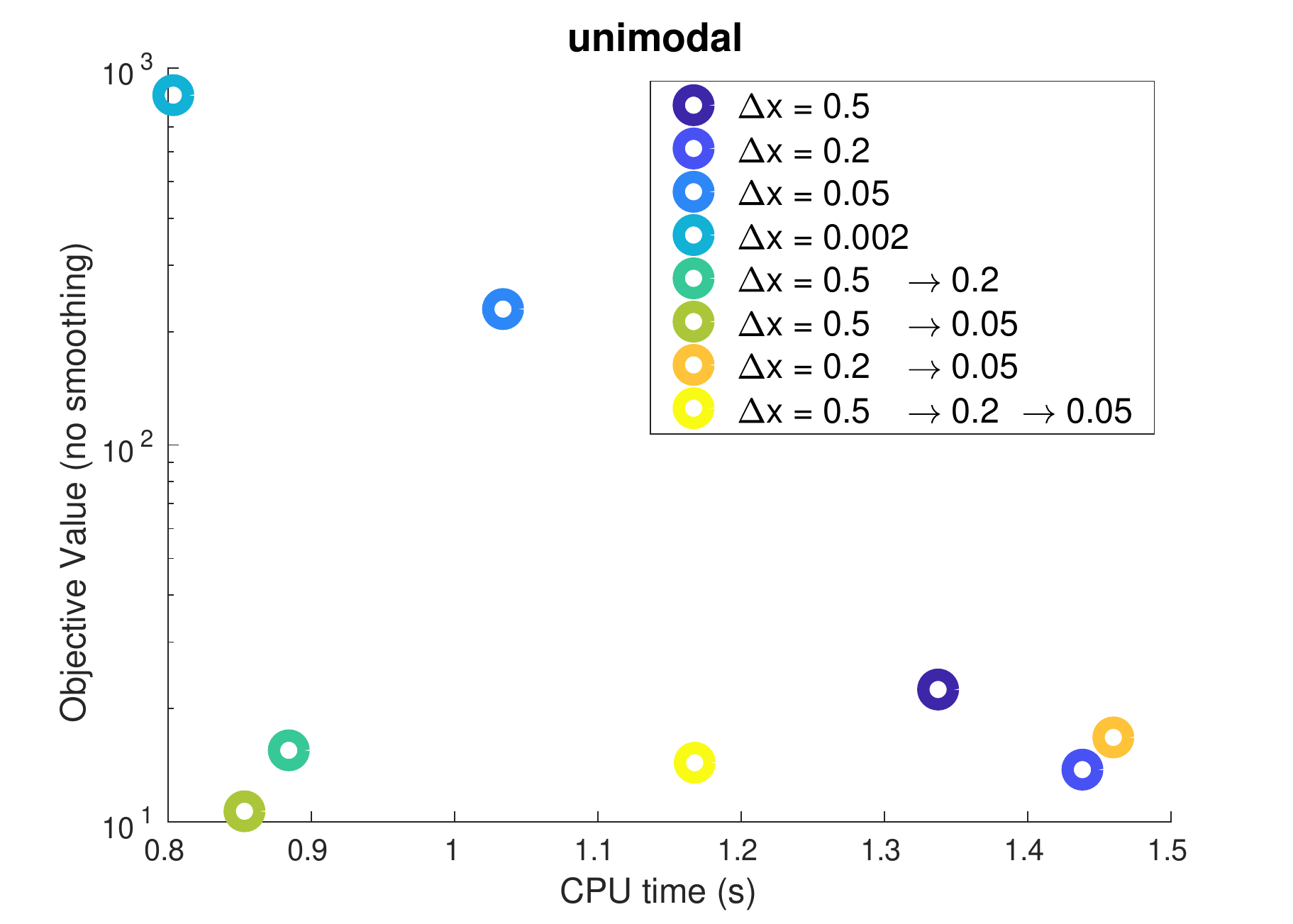}
        \caption{}
    \end{subfigure}%
    ~ 
    \begin{subfigure}[b]{0.5\textwidth}
        \centering
        \includegraphics[width=\textwidth]{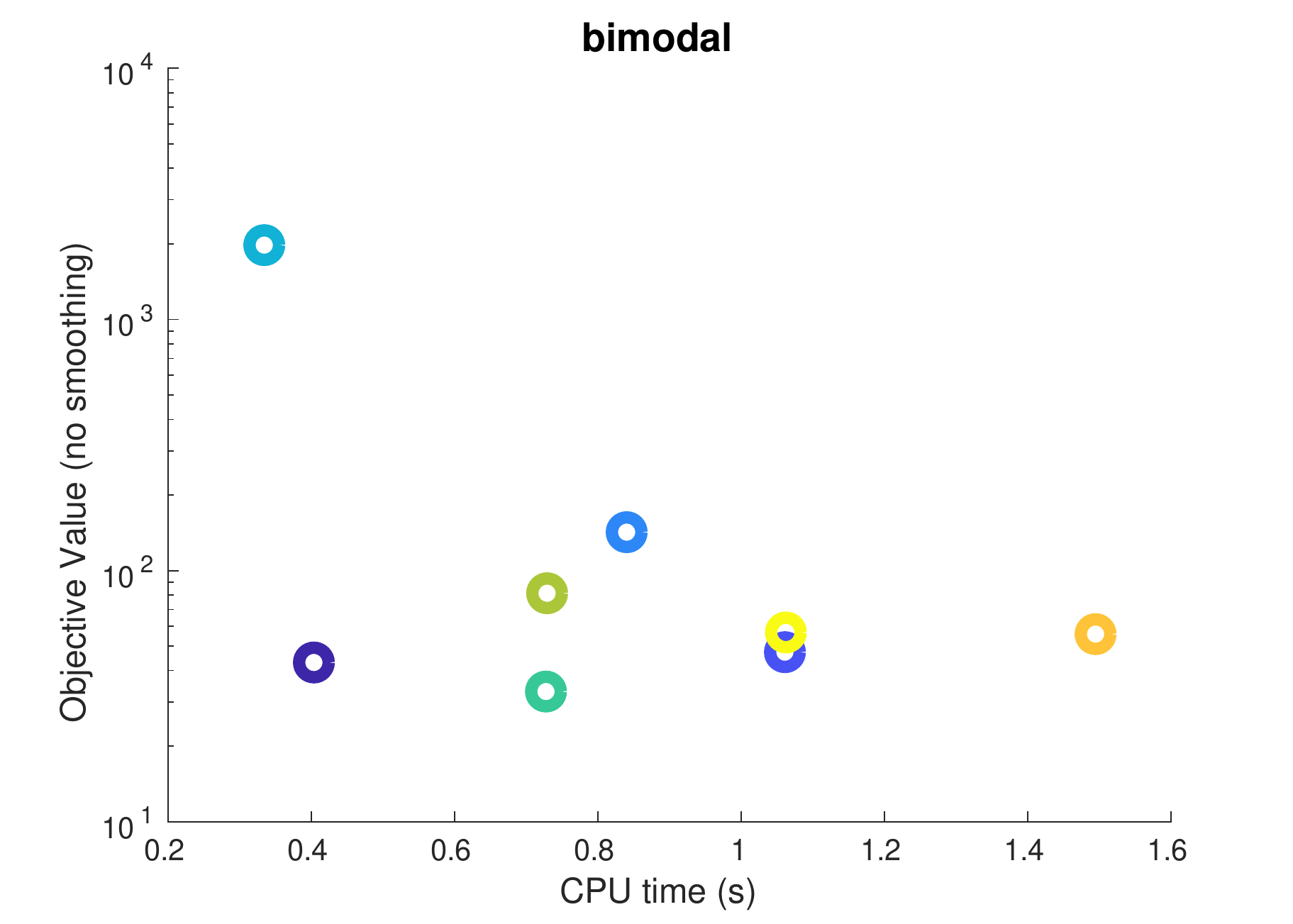}
        \caption{}
    \end{subfigure}
    \caption{Comparison of smoothing parameter schemes. In each case, the smoothing in the legend is used during the optimization, but the objective values shown on the plot are computed without the smoothing function. This provides a uniform comparison for all trials. The horizontal axis shows the CPU time to compute each solution, for both the unimodal (left panel) and bimodal (right panel) fluence profiles using $T=5$ seconds of delivery. Note the $\Delta x =0.002$ case is effectively ``no smoothing''. Note the right arrows in the smoothing schemes legend refer to discrete transitions of the smoothing parameters.}
  \label{fig:smoothingParamSweep_pareto}
\end{figure*}

Before the algorithm can be run, a smoothing parameter scheme -- a sequence of smoothing parameter values $\alpha$ -- has to be chosen.
Each smoothing parameter value is derived from $\gamma$ and $\Delta x$ using equation (\ref{eqn:SmoothingDistanceParameter}) and holds during a certain stage of the algorithm. 
We choose to use a constant $\gamma = 0.95$ and study three values for the smoothing distance $\Delta x = \{0.5, 0.2, 0.05\}$ cm.
We explore each possible sequence of smoothing stages for these three parameters for which the level of smoothing decreases, that is, the accuracy increases (see the legend of Figure \ref{fig:smoothingParamSweep_pareto}).
We also include an additional trial with a width of $\Delta x = 0.002$, which is effectively equivalent to no smoothing.
Every smoothing stage is solved using the $\mathsf{fmincon}$ function in Matlab Release R2017a (The MathWorks, Inc, Natick, USA) with default settings.

The algorithm progresses from one smoothing stage to the next (or terminates if the current stage is the last stage) when $\mathsf{fmincon}$ converges. 
The only difference between the optimization in two consecutive stages is the smoothing distance $\Delta x$ and the initialization.
The algorithm is initialized with the leaf pair moving at a constant speed, sweeping the entire domain with a fixed small leaf gap. Subsequent stages use the solution from the previous stage as initialization.

Figure \ref{fig:smoothingParamSweep_pareto} shows the optimization results for each of the smoothing parameter schemes, represented by the objective value of the final solution and CPU time, under a moderate delivery time of $T=$ 5 seconds (using 11 knot points). For comparison of the solutions, the objective value is computed without smoothing. Heavy smoothing (large $\Delta x$) results in fast optimization but poor fitting, whereas light smoothing results in slow optimization as well as poor fitting.
The best solutions were obtained by starting with heavy smoothing and then moving to moderate smoothing.
These solutions require a moderate amount of CPU time but tend to be more accurate than most other methods.
We use the $ \Delta x = 0.5 \rightarrow 0.2 $ smoothing scheme in subsequent experiments.

Note that using negligible smoothing ($ \Delta x=0.002$ cm) yields poor results, as the optimization fails to converge. This indicates the importance of smoothing and iterative refinement of the smoothing parameter.

\subsection{Progress of the Algorithm}
\begin{figure}
  \centering
  \includegraphics[width=\textwidth]{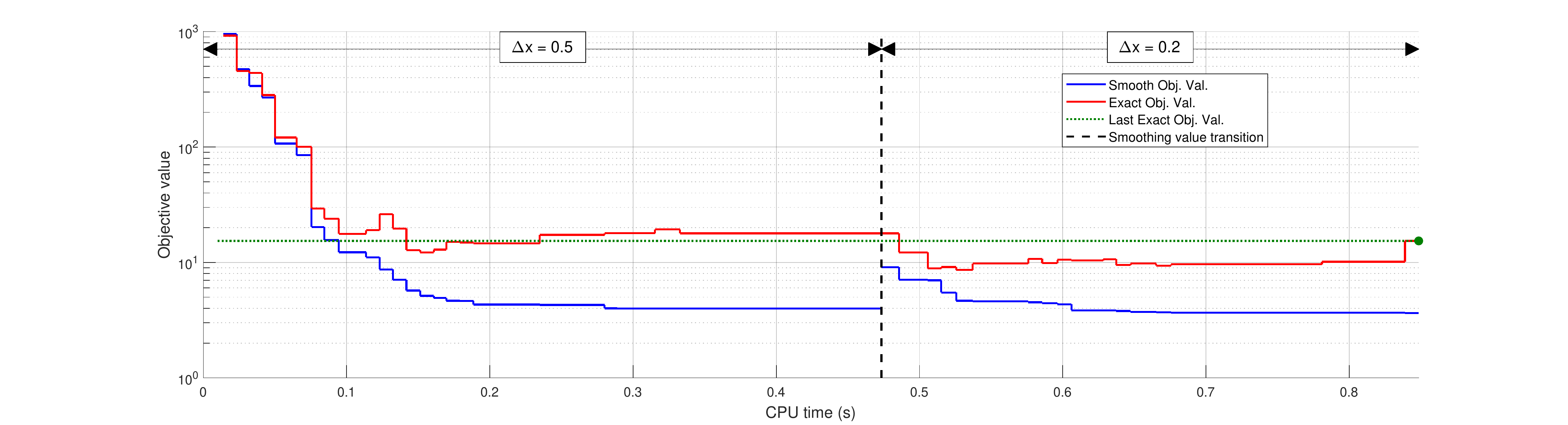}
  \caption{Relation between CPU time and the best known smoothed objective value and the corresponding exact objective value, for the parameter scheme ($\Delta x= 0.5 \to 0.2$) and the unimodal fluence profile (see Figure \ref{fig:smoothingParamSweep_unimodalTraj}), with $T=5$s.}
  \label{fig:CPUvsObj_bestalpha_uni}
\end{figure}

\begin{figure}
  \centering
  \includegraphics[width=\textwidth]{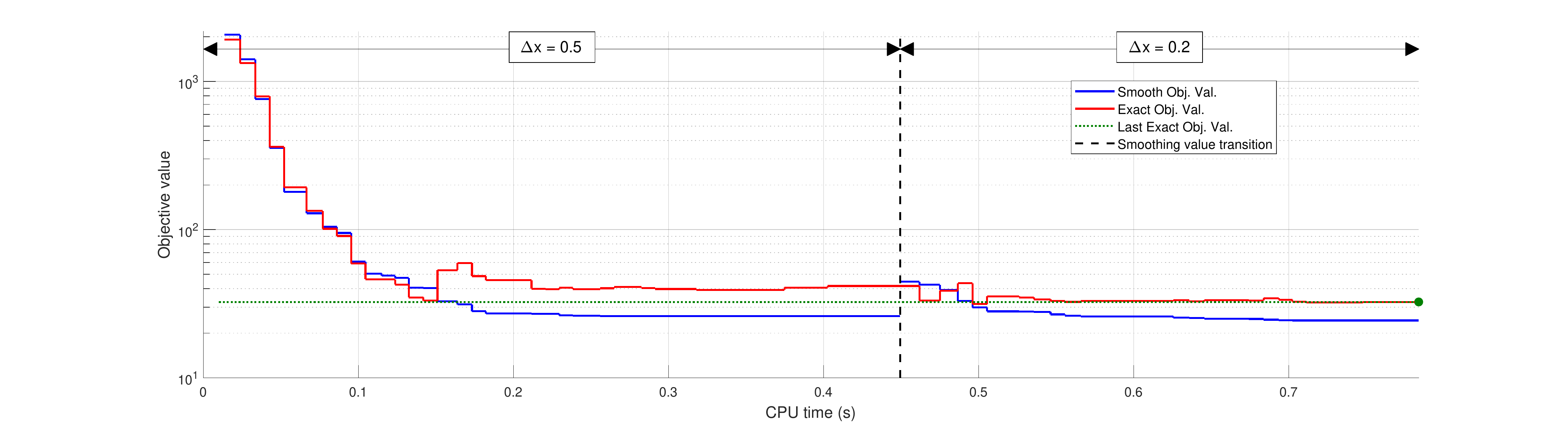}
  \caption{Relation between CPU time and the best known smoothed objective value and the corresponding exact objective value, for the parameter scheme ($\Delta x= 0.5 \to 0.2$) and the bimodal fluence profile (see Figure \ref{fig:smoothingParamSweep_bimodalTraj}), with $T=5$s.}
  \label{fig:CPUvsObj_bestalpha_bi}
\end{figure}

For the $\Delta x = 0.5 \rightarrow 0.2$ smoothing scheme and $T=5$ seconds of delivery, Figures \ref{fig:CPUvsObj_bestalpha_uni} and \ref{fig:CPUvsObj_bestalpha_bi} show the quality of the current solution as the algorithm proceeds, for the unimodal and bimodal case, respectively. For both the unimodal and bimodal case a decent solution is found halfway through the first smoothing stage, which is then fine-tuned as the algorithm proceeds. 

By definition, the objective value of the best known solution - evaluated at the currently active smoothing level (blue line) - is decreasing in CPU time within every stage of the algorithm.
At the start of a new stage, an improvement in the smoothed objective value typically results in an improvement in the exact objective value as well.
Later on, when improvements in the smoothed objective value are smaller, the corresponding effect on the exact objective value can be of either sign but is generally small. 
Naturally, when transitioning from one smoothing stage to another, the smoothed objective value instantly changes whereas the corresponding exact objective value is unaltered.

In general it is not guaranteed that the exact objective value of the last found solution (green line) is the best solution.
Therefore, we not only keep track of the current solution and its objective value, but also track the solution with the best exact objective value, and at termination accept the latter as our final solution. We also note that the best solutions found by our algorithm, even when $T$ is sufficient for perfect matching, do not show an objective value of 0. This is related to how trajectories are represented, to the level of discretization, and to how fluence maps are computed from trajectories. The inter-dependencies of these modeling choices are discussed in \cite{thesisKvA}, but we note that the plateau objective function values, while not numerically 0, are sufficiently small such that the recreated fluence maps are visually indistinguishable from the target maps (see Figures \ref{fig:TvsObj_bestalpha_uni} and \ref{fig:TvsObj_bestalpha_bi}, rightmost solutions).

\subsection{Leaf Trajectories}
\begin{figure}[ht]
  \centering
  \begin{subfigure}[b]{0.5\linewidth}
    \centering\includegraphics[width=0.9\linewidth]{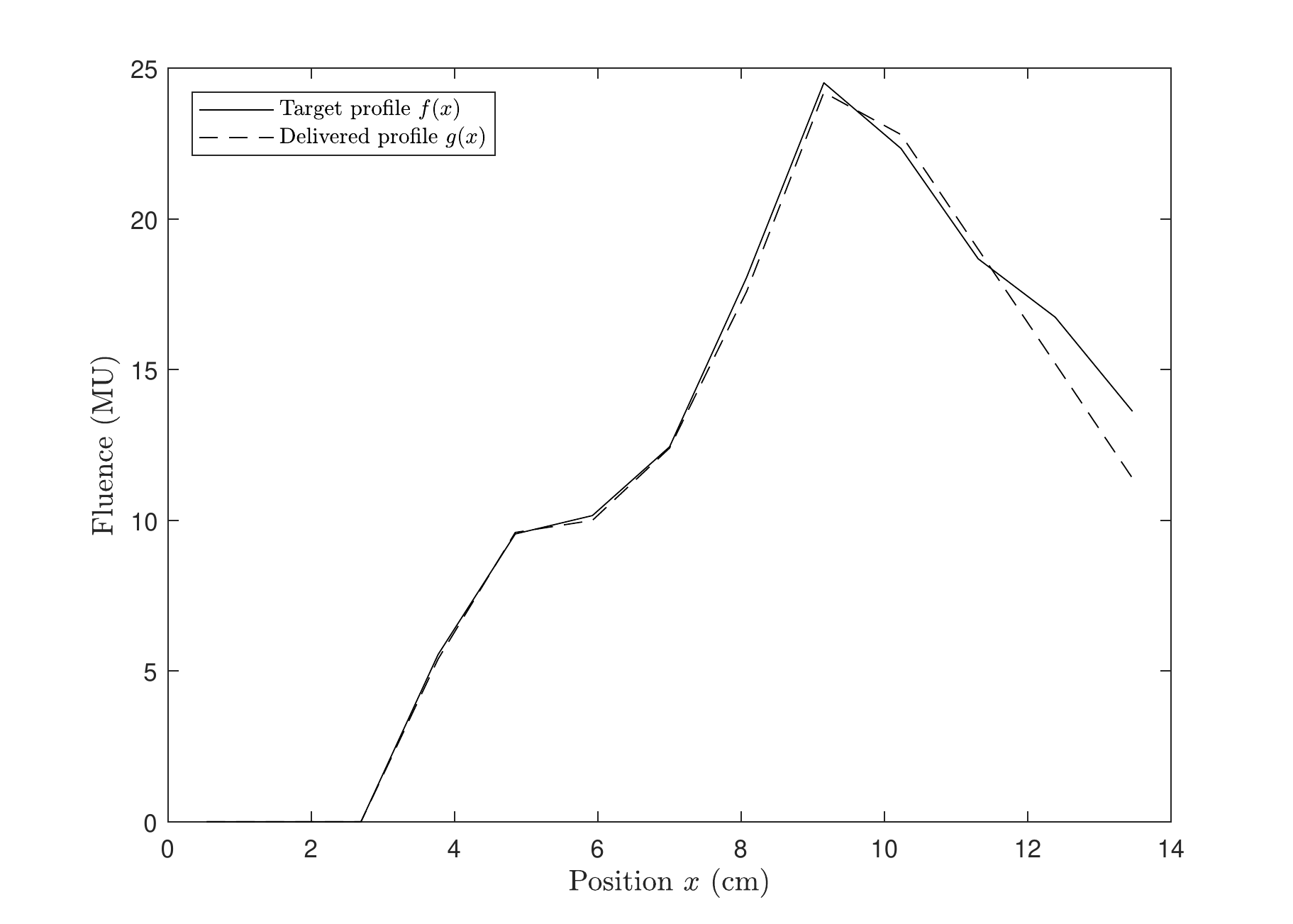}
    \caption{\label{fig:sps_umt_prof}}
  \end{subfigure}%
  \begin{subfigure}[b]{0.5\linewidth}
    \centering\includegraphics[width=0.9\linewidth]{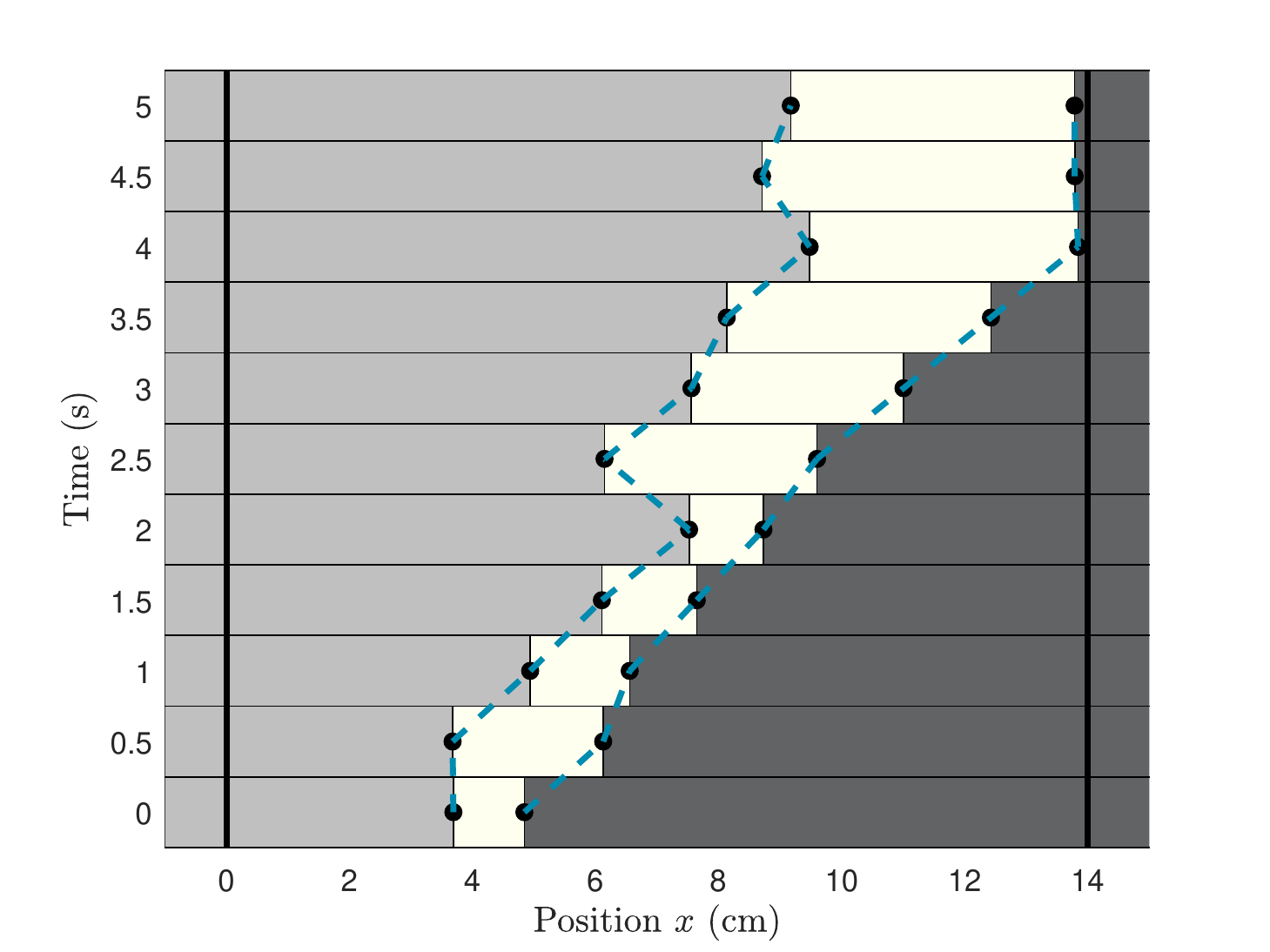}
    \caption{\label{fig:sps_umt_traj}}
  \end{subfigure}
  \caption{Within $T=5$ seconds of delivery, the unimodal target profile (solid line, left panel) is closely matched (dashed line, left panel) by the leaf trajectories displayed in the right panel. These trajectories are found by optimization using the ($\Delta x = 0.5 \to 0.2$) smoothing schedule.}
  \label{fig:smoothingParamSweep_unimodalTraj}
\end{figure}

\begin{figure}[ht]
  \centering
  \begin{subfigure}[b]{0.5\linewidth}
    \centering\includegraphics[width=0.9\linewidth]{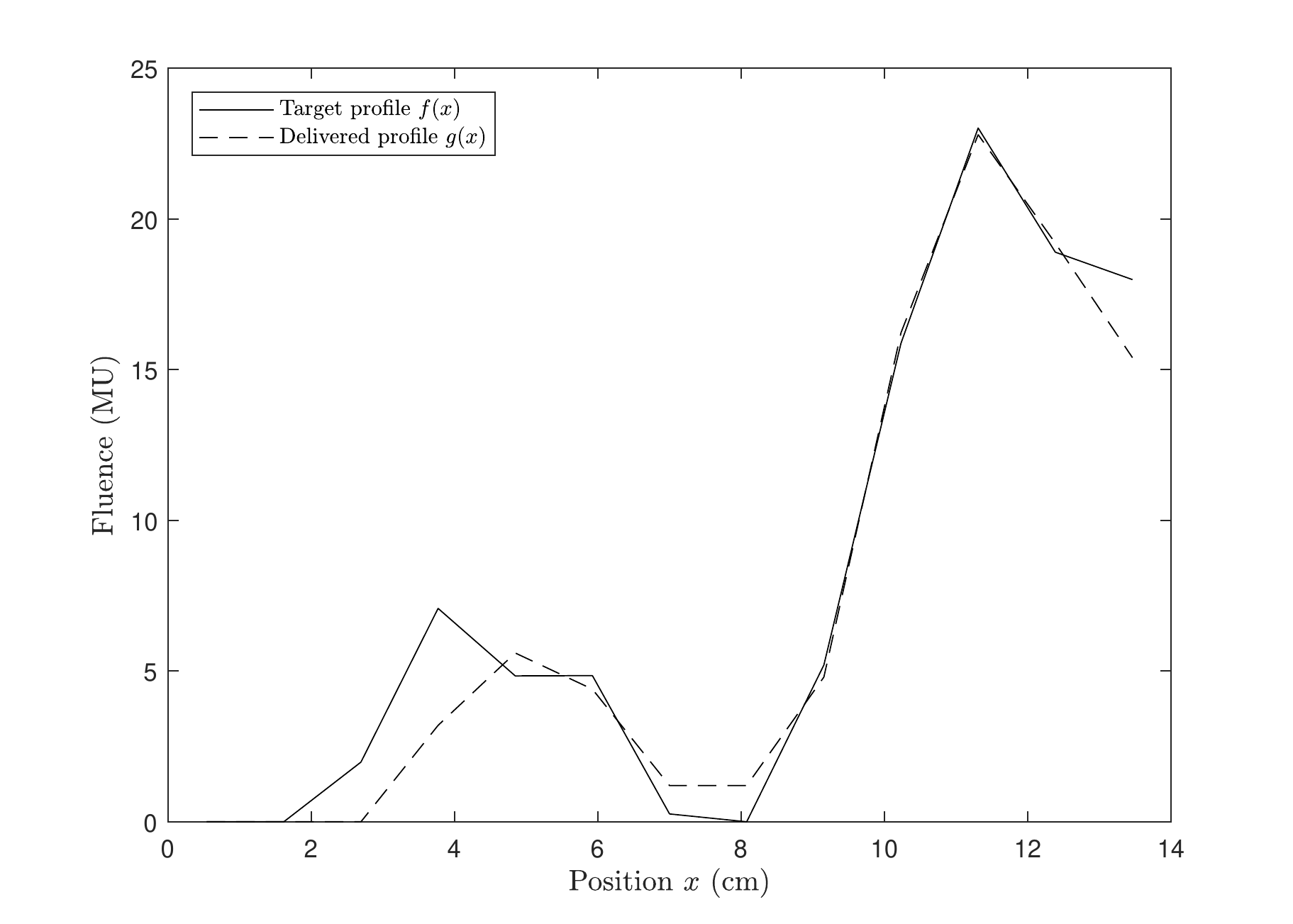}
    \caption{\label{fig:sps_bmt_prof}}
  \end{subfigure}%
  \begin{subfigure}[b]{0.5\linewidth}
    \centering\includegraphics[width=0.9\linewidth]{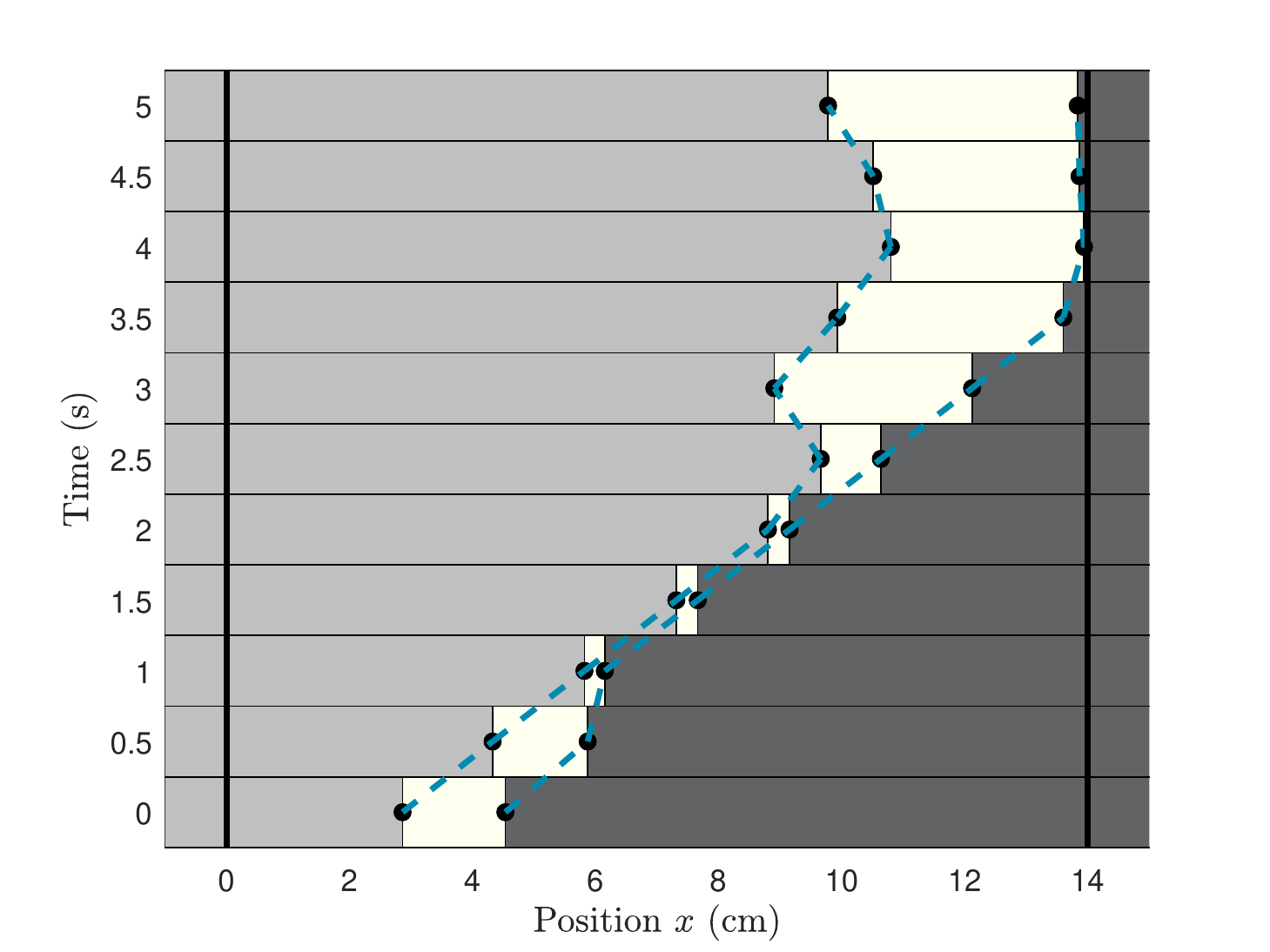}
    \caption{\label{fig:sps_bmt_traj}}
  \end{subfigure}
  \caption{Within $T=5$ seconds of delivery, the bimodal target profile (solid line, left panel) is well matched (dashed line, left panel) by the leaf trajectories displayed in the right panel. These trajectories are found by optimization using the ($\Delta x = 0.5 \to 0.2$) smoothing schedule.}
  \label{fig:smoothingParamSweep_bimodalTraj}
\end{figure}

Figures \ref{fig:smoothingParamSweep_unimodalTraj} and \ref{fig:smoothingParamSweep_bimodalTraj} show the fluence profile (left panel) delivered by the leaf trajectories of the final solution (right panel) for the unimodal and bimodal case, respectively.
In the unimodal case, the targeted fluence profile is almost perfectly matched, using a final solution in which leaves move in a near-unidirectional fashion.
In fact, we could swap the position of the left leaf at $T=2.5$s with the position of that leaf at $T=3$ and move that of $T=4$s to the end without changing the delivered fluence profile while respecting constraints.
In fact, if the dose rate is constant, every pair of non-unidirectional leaf trajectories can be transformed into a pair of unidirectional leaf trajectories, without changing the delivered fluence profile, as shown in the Appendix of \cite{balvertcraft}.
This illustrates that there might be multiple optimal solutions to our problem.
Note that perfect delivery could be achieved with leaves moving in a unidirectional fashion if the delivery time would be larger than or equal to the SWLS row delivery time (5.8s for this fluence row; 6.7s for the entire map, see Figure \ref{fig:TvsObj_bestalpha_uni}).

In the bimodal case the matching is not as close but still reasonably good. This is because the available delivery time (5s) is smaller than the SWLS row delivery time for this profile (6.7s).
Again, the leaves move in a near-unidirectional like fashion, but could as well have moved fully unidirectionally. 
Mismatches occur at the boundaries of the field and at the dip in the fluence profile.
In order to modulate the dip in the fluence profile, the leaves would have to fully close, by which the leaves would, with restricted time as is, not be able to modulate other parts, for which the price of not spending sufficient time there is higher.
Naturally, the bounds of the field are harder to deliver as there is less flexibility in how and when to expose these parts to radiation. 

\subsection{Matching an Entire Fluence Map}
In clinical practice one always faces the challenge of matching an entire fluence map rather than only a single row. An upper bound on the time needed to perfectly match a fluence map is the maximum of the SWLS row delivery time over all rows. For the near-unimodal map depicted in Figure \ref{fig:targetmap} and its transposed near-bimodal version, these are 6.7 s and 8.6 s respectively.

Keeping the dose rate fixed to its maximum level, we utilize the independence property to optimize the leaf trajectories of every single leaf pair in parallel. By running these optimizations for all integer delivery times $T$ between one and the upwards rounded SWLS delivery time, the trade-off between delivery time and fluence map matching quality is generated. 

\begin{figure}
  \centering
  \includegraphics[width=\textwidth]{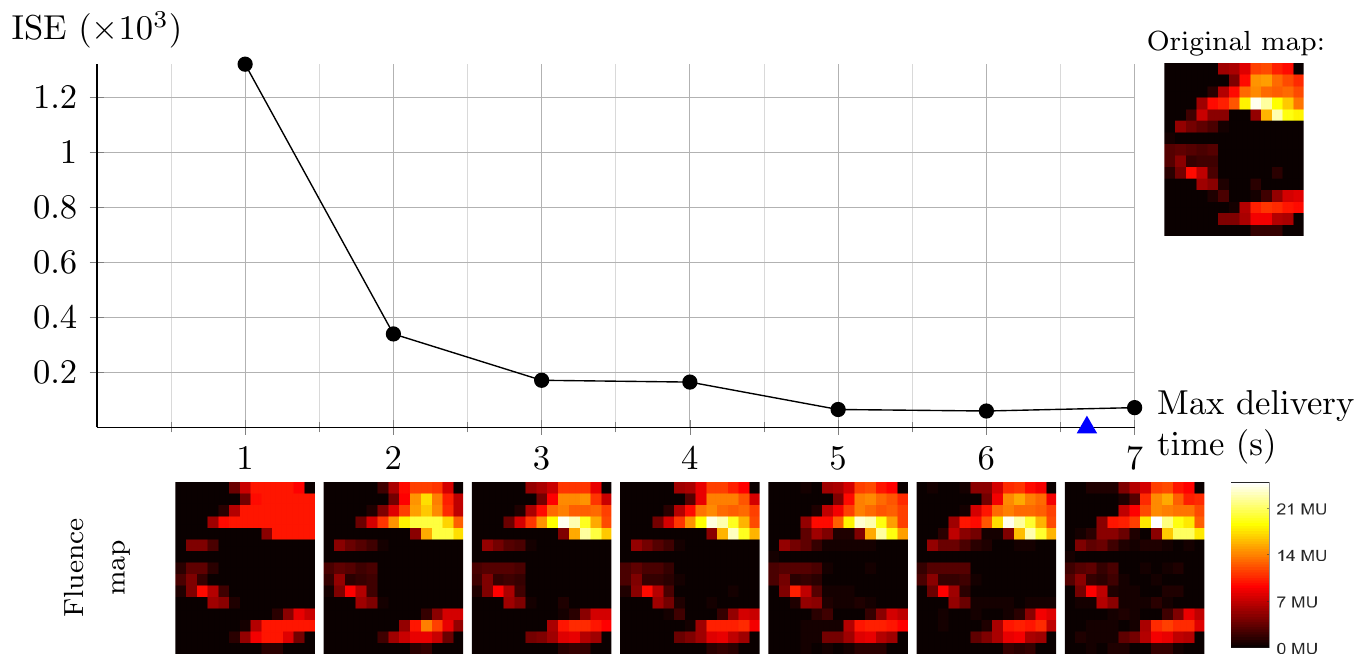}
  \caption{Solution to the problem of matching the entire fluence map (see Figure \ref{fig:targetmap}), for various delivery times. The vertical axis of the graph shows the exact objective function, the integral squared error (ISE), that is minimized. For each integer second of delivery time, the delivered fluence map is shown. The blue triangle represents the SWLS time (6.7s).}
  \label{fig:TvsObj_bestalpha_uni}
\end{figure}

For the near-unimodal fluence map studied, this trade-off curve is depicted in Figure \ref{fig:TvsObj_bestalpha_uni}. With just a single second of delivery, the contours of the map are largely delivered. When more time becomes available, the delivery window concentrates more on the fluence peaks.
One can achieve near-perfect fluence map matching within 5 seconds. With 3 or 4 seconds of delivery, the degradation in solution quality is minor. For larger delivery times, there is no improvement in the sum of squared errors. As the number of variables and hence the number of dimensions in the solution space is increasing with delivery time, this is likely caused by the algorithm getting stuck in a local optimum. 

\begin{figure}
  \centering
  \includegraphics[width=\textwidth]{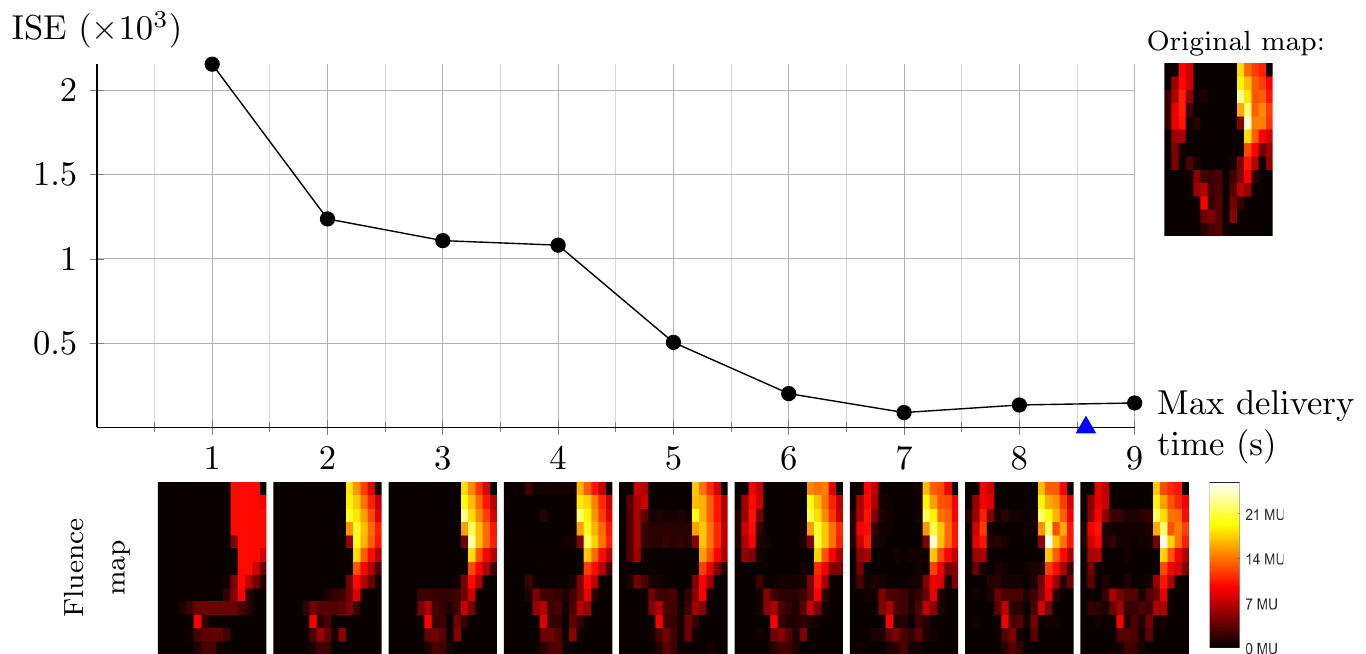}
  \caption{Solution to the problem of matching the entire, transposed fluence map (see Figure \ref{fig:targetmap}), for various delivery times. The vertical axis of the graph shows the exact objective function, the integral squared error (ISE), that is minimized. For each integer second of delivery time, the delivered fluence map is shown. The blue triangle represents the SWLS time (8.6s).}
  \label{fig:TvsObj_bestalpha_bi}
\end{figure}

For the bimodal case, Figure \ref{fig:TvsObj_bestalpha_bi}, more delivery time is needed to achieve decent matching: it takes 5 seconds in the unimodal case to drop below an ISE of 0.5$\times 10^3$, versus 7 seconds for the bimodal case. In the bimodal case, since exposing the whole width of the field would do too much harm to the untargeted center segment, the leaves first focus on the most intense half of the map. As 2s or 3s is insufficient to cross the field, for those delivery times the right peak matching is improved upon rather than trying to deliver the left peak as well. When the leaves can make it across the gap (4-5s), the left peak is modulated, at the price of losing precision in the right peak, but with the merit of quick overall matching improvement. The time required to modulate the right part is now needed to traverse the field. For even longer delivery times, delivery at both sides is smoothened. 

For both bimodal and unimodal cases, near-optimal matching is possible with a delivery time that is significantly smaller (on the order of 20-25\%) than the SWLS delivery time.

\section{Discussion and conclusions}

The pure fluence map sequencing problem has received little attention in the past few years compared to full VMAT optimization. VMAT presents a clinically relevant and algorithmically challenging problem, but since dynamic fluence map sequencing lies at the heart of VMAT optimization (even though few optimization approaches recognize this), we are interested in returning to that basic, unsolved problem.

We begin by assuming that fluence maps are given as the result of an optimization procedure. The community of radiotherapy optimization researchers also needs to continually consider the dose optimization problem, a step that we do not address in this report. This step is of fundamental importance, where the challenge is not as much in optimization algorithms but knowing what to optimize. Target definition and dose tolerance and prescription levels are set based on historical knowledge rather than biologically informed criteria. Although one could argue that this aspect of treatment planning research is of more fundamental importance, the community will need to improve on both 'what to optimize' and 'how to best deliver the optimized fluence maps' in order to advance patient care, and we address the second area herein. 

The dynamic fluence map sequencing problem has been visited before in \cite{balvertcraft} (and references therein). Both their procedure and ours model the leaves and dose rate by specifying their values at several moments in time (the ``knot points'' in our model). The main difference lies in the formulation of the exposure function: while we use a continuous function, like \cite{pappvmat}, the exposure in \cite{balvertcraft}, which follows the more common way that IMRT and VMAT are modeled, is based on a discretized approximation of the inherently continuous leaf trajectories. This makes our approach more realistic than the method used in \cite{balvertcraft}. Comparing the two methods in terms of fluence map replication accuracy is complicated by the difference in model formulation: a continuous exposure function asks for a continuous objective value, namely ISE, whereas the use of a discretized exposure function requires the user to evaluate plan quality with the sum of squared errors (ssdif). If one were to evaluate both procedures with ISE, then our proposed fully continuous method would come out as the best performer, and vice versa. Although ssdif is essentially a discretization of ISE, their values may differ significantly, particularly because ssdif is likely to be an underestimation of the true delivery error. It is therefore difficult to compare the fluence map replication accuracy of the two methods. However, the shape of the trade-off curve in Figure \ref{fig:TvsObj_bestalpha_uni} is very similar to its ssdif equivalent in Figure 4 in \cite{balvertcraft}. This indicates that both methods perform in a similar manner, with ours considering more realistic (i.e. continuous rather than discretized) representations of the leaf trajectories and fluence computations.

The fundamental difficulty in the single map dynamic sequencing problem (and in turn, VMAT) is nonconvexity, which rears its head in the many local optima of the objective function, a large number of which are comparable to the global optimum. For VMAT the large number of local optimal solutions of similar high quality can be loosely justified by noting that in the case of coplanar IMRT, a large number of equispaced beams (say, 15 or more, see \cite{bortBeams}) will provide an optimal solution independent of their exact location. Thus there is freedom in the start and end gantry angles for delivery of the individual fields. This implies many optimal solutions, since it is likely that good leaf positions could be almost anywhere within the bounds of the field at any given gantry position. Still, finding and verifying any one of these good local optima is not an easy task, which is a direct consequence of the near impossibility of obtaining certificates of global optimality for nonconvex optimization problems. Mixed integer linear programming formulations offer a possible approach here \cite{mipvmat1,mipvmat2,mipvmat3}, but the challenges of formulating the problem with continuous fluence computations and variable gantry speed and dose rates, along with the formidable computational expense of solving such problems, have kept such approaches away from clinical usage thus far.

When problems have many near optimal solutions, it makes sense to regularize the search space, which in our case can be done by restricting needless back and forth motions of the MLC leaves. One way to accomplish this, which to our knowledge has not been studied before, is to represent leaf trajectories with reduced degrees of freedom. This could be done by a coarse discretization of the leaf position versus time space, or piecewise linear leaf trajectories (piecewise constant leaf speeds), which yields the benefit of a coarser trajectory description while retaining an accurate leaf position versus time description for fluence transmission computations.

There are other choices for representing trajectories, most of which would fall into the category of polynomial splines. There is a fundamental trade-off in polynomial splines: for a given amount of data you can store many low-order segments or few high-order segments. Selecting the correct trade-off is discussed in detail in \cite{kelly2017introduction}, \cite{Betts2010}, \cite{Darby2011a}. 

One reason for our choice of linear splines is that we can precisely enforce velocity constraints without the need for mesh refinement or other expensive checks. The low order spline also lends itself to fast and simple calculations. Finally, linac control systems themselves use linear interpolation between control points. We performed a brief pilot study evaluating linear versus cubic splines, with the same number of decision variables in the optimization. We found that the linear splines resulted in faster optimization for a comparable accuracy and dramatically simplified the resulting optimization code.

In addition to the spline representation, we also use a controllable smoothing function to smooth the typical step function representation of an MLC leaf blocking radiation. While we introduced it for its numerical benefits, it is also a more realistic model of a leaf blocking radiation: due to leaf tip scatter, the fluence will never be a sharp step function. The standard technique of beginning with a large amount of smoothing and gradually decreasing it worked well, although in general this smoothing schedule could be automated and optimized.

If one had an algorithm that, given a dose rate profile and a delivery time limit $T$, returned optimal leaf trajectories, one could then build an outer loop algorithm that searched the dose rate profile space. One could also represent the continuous dose rate as a piecewise linear spline, to regularize that search as well. Due to the decoupling of the MLC leaf rows, we envision that this is a worthwhile way to pursue the entire problem. Global techniques which find a balance between exploration and exploitation, such as Bayesian Optimization or CMAES \cite{Hansen2001}, are possible dose rate search strategies. We recommend searching a parameterized dose rate profile space that ``makes sense'', rather than blindly searching across all feasible dose rate profiles. For example, one generally wants the dose rate to be at its maximal value, with occasional dips (possibly to 0) to allow leaves to reposition without delivering dose. Dose rate search is a difficult problem however, and warrants a full investigation. Finally, it remains to be seen if this nested approach (outer loop dose rate, inner loop leaf trajectories) should be pursued or replaced by a different search style. If nested optimization is pursued, details including how many inner loop iterations for a given outer loop dose setting need to be explored.


Mathematically it is straightforward to extend these ideas to the case of full VMAT optimization. However, additional decision variables for gantry speed, and in general the much larger number of control points needed, would make such an approach computationally infeasible. We thus consider optimal VMAT optimization very much an open question.

\bigskip

\bibliographystyle{unsrt}
\bibliography{all}

\end{document}